\documentclass[a4paper]{jpconf}

\begin{document}

\title{Dirac fermions and Kondo effect
}

\author{Takashi \textsc{Yanagisawa}
}

\address{Electronics and Photonics Research Institute,
National Institute of Advanced Industrial Science and Technology (AIST),
Tsukuba Central 2, 1-1-1 Umezono, Tsukuba 305-8568, Japan
}

\ead{t-yanagisawa@aist.go.jp}


\begin{abstract}
In this study, we investigate the Kondo effect induced by the s-d interaction 
where the conduction bands are
occupied by Dirac fermions.  The Dirac fermion has the linear dispersion and 
is described typically
by the Hamiltonian such as $H_k= v{\bf k}\cdot {\sigma}+m \sigma_0$ for the 
wave number ${\bf k}$ 
where $\sigma_j$ are
Pauli matrices and $\sigma_0$ is the unit matrix.  We derived the formula of 
the Kondo temperature $T_{\rm K}$ by means of 
the Green's function theory for Green's functions including Dirac fermions and 
the localized spin.  
The $T_{\rm K}$ was determined from a singularity of Green's functions in the form
$T_{\rm K}\propto \exp(-{\rm const}/\rho|J|)$.
The Kondo effect will disappear when the Fermi surface is point like because
$T_{\rm K}$ vanishes as the chemical potential
$\mu$ approaches the Fermi point.
\end{abstract}
 


\section{Introduction}

Recently, the Dirac electron in solid state has been investigated 
intensively [1-6].
It is interesting to examine how the Kondo effect occurs in a system of 
Dirac electrons with magnetic impurities.  
It is not so trivial whether the Kondo effect indeed appears there.
We expect significant and interesting behaviors when the localized spin 
interacts with the Dirac electron through the
s-d interaction.  The Dirac Hamiltonian resembles the s-d model with the 
spin-orbit coupling of Rashba type [7].
In this paper we investigate the s-d Hamiltonian with Dirac electrons by 
means of the Green's function theory and 
evaluate the Kondo temperature $T_{\rm K}$.

The Hamiltonian is given by $H=H_0+H_{sd}\equiv H_m+H_K+H_{sd}$ where
\begin{eqnarray}
H_m &=& \sum_{{\bf k}}(m-\mu)(c_{{\bf k}\uparrow}^{\dag}c_{{\bf k}\uparrow}
+c_{{\bf k}\downarrow}^{\dag}c_{{\bf k}\downarrow}),\\
H_K&=& \sum_{{\bf k}}[ v(k_x-ik_y)c_{{\bf k}\uparrow}^{\dag}
c_{{\bf k}\downarrow}
+v(k_x+ik_y)c_{{\bf k}\downarrow}^{\dag}c_{{\bf k}\uparrow}\nonumber\\
&+&v_zk_z(c_{{\bf k}\uparrow}^{\dag}c_{{\bf k}\uparrow}
-c_{{\bf k}\downarrow}^{\dag}c_{{\bf k}\downarrow})]\\
H_{sd}&=& -\frac{J}{2}\frac{1}{N}\sum_{{\bf k}{\bf k}'}[
S_z(c_{{\bf k}\uparrow}^{\dag}c_{{\bf k}'\uparrow}-
c_{{\bf k}\downarrow}^{\dag}c_{{\bf k}'\downarrow})
+S_{+}c_{{\bf k}\downarrow}^{\dag}c_{{\bf k}'\uparrow}
+S_{-}c_{{\bf k}\uparrow}^{\dag}c_{{\bf k}'\downarrow}].
\end{eqnarray}
$v$ and $v_z$ are velocities of conduction electrons, $\mu$ is the chemical
potential and $m$ is the mass of the Dirac fermion.
$N$ indicates the number of sites.
We set $m=0$ so that the chemical potential $\mu$ includes $m$.
$c_{{{\bf k}}\sigma}$ and $c_{{{\bf k}}\sigma}^{\dag}$ are annihilation and
creation operators, respectively.
$S_{+}$, $S_{-}$ and $S_z$ denote the operators of the localized spin.
The term $H_{sd}$ indicates the s-d interaction between the conduction electrons
and the localized spin, with the coupling constant $J$ \cite{kon64,kondo}.
$J$ is negative, as adopted in this paper, for the antiferromagnetic interaction.

\section{Green's Functions}

We define thermal Green's functions of the conduction electrons
\begin{eqnarray}
G_{{\bf k}{\bf k}'\sigma}(\tau)&=&-\langle T_{\tau}c_{{\bf k}\sigma}(\tau)
c_{{\bf k}'\sigma}^{\dag}(0)\rangle,\\
F_{{{\bf k}}{{\bf k}}'}(\tau)&=&-\langle T_{\tau}c_{{{\bf k}}\downarrow}(\tau)
c_{{{\bf k}}'\uparrow}^{\dag}(0)\rangle,
\end{eqnarray}
where $T_{\tau}$ is the time ordering operator.
We note that the spin operators satisfy the following relations:
\begin{eqnarray}
S_{\pm}S_z&=&\mp\frac{1}{2}S_z,~~~S_zS_{\pm}=\pm\frac{1}{2}S_{\pm},\\
S_+S_-&=&\frac{3}{4}+S_z-S_z^2,\\
S_-S_+&=&\frac{3}{4}-S_z-S_z^2.
\end{eqnarray}

We also define Green's functions which include the localized spins 
as well as the
conduction electron operators.   They are for example, following the 
notation of
Zubarev \cite{zub60},
\begin{eqnarray}
\langle\langle S_zc_{{{\bf k}}\uparrow};c_{{{\bf k}}'\uparrow}^{\dag}
\rangle\rangle_{\tau}
&=& -\langle T_{\tau}S_z c_{{{\bf k}}\uparrow}(\tau)c_{{{\bf k}}'\uparrow}^{\dag}(0)\rangle,\\
\langle\langle S_{-}c_{{{\bf k}}\downarrow};c_{{{\bf k}}'\uparrow}^{\dag}
\rangle\rangle_{\tau}
&=& -\langle T_{\tau}S_{-}c_{{{\bf k}}\downarrow}(\tau)c_{{{\bf k}}'\uparrow}^{\dag}(0)
\rangle,\\
\langle\langle S_zc_{{{\bf k}}\downarrow};c_{{{\bf k}}'\uparrow}^{\dag}\rangle\rangle_{\tau}
&=& -\langle T_{\tau}S_z c_{{{\bf k}}\downarrow}(\tau)c_{{{\bf k}}'\uparrow}^{\dag}(0)
\rangle,\\
\langle\langle S_-c_{{{\bf k}}\uparrow};c_{{{\bf k}}'\uparrow}^{\dag}
\rangle\rangle_{\tau}
&=& -\langle T_{\tau}S_- c_{{{\bf k}}\uparrow}(\tau)c_{{{\bf k}}'\uparrow}^{\dag}(0)
\rangle.
\end{eqnarray}
The Fourier transforms are defined as usual:
\begin{eqnarray}
G_{{{\bf k}}{{\bf k}}'\sigma}(\tau)&=& \frac{1}{\beta}\sum_n e^{-i\omega_n\tau}
G_{{{\bf k}}{{\bf k}}'\sigma}(i\omega_n),\\
F_{{{\bf k}}{{\bf k}}'}(\tau)&=& \frac{1}{\beta}\sum_n e^{-i\omega_n\tau}
F_{{{\bf k}}{{\bf k}}'}(i\omega_n),\\
\langle\langle S_zc_{{{\bf k}}\uparrow};c_{{{\bf k}}'\uparrow}^{\dag}
\rangle\rangle_{\tau}&=&
\frac{1}{\beta}\sum_n e^{-i\omega_n\tau}
\langle\langle S_zc_{{{\bf k}}\uparrow};c_{{{\bf k}}'\uparrow}^{\dag}
\rangle\rangle_{i\omega_n},\\
&& \cdots\cdots.\nonumber
\end{eqnarray}

Using the commutation relations
the equation of motion for $G_{{{\bf k}}{{\bf k}}'\uparrow}(i\omega_n)$ reads
\begin{eqnarray}
(i\omega_n-v_z k_z-m+\mu )G_{{\bf k}{\bf k}'\uparrow}(\omega_n)&=&
\delta_{{\bf k}{\bf k}'}
-\frac{J}{2N}\sum_{q}
\Gamma_{{\bf q}{\bf k}'} (i\omega_n)\nonumber\\
&+& v(k_x-ik_y)F_{{\bf k}{\bf k}'} (i\omega_n).
\end{eqnarray}
Here we have defined
\begin{eqnarray}
\Gamma_{{{\bf k}}{{\bf k}}'}(\tau)&=&\frac{1}{\beta}\sum_n e^{-i\omega_n}
\Gamma_{{{\bf k}}{{\bf k}}'}(i\omega_n)
= \langle\langle S_zc_{{{\bf k}}\uparrow};c_{{{\bf k}}'\uparrow}^{\dag}
\rangle\rangle_{\tau}
+ \langle\langle S_-c_{{{\bf k}}\downarrow};c_{{{\bf k}}'\uparrow}^{\dag}
\rangle\rangle_{\tau}.
\end{eqnarray}
The equation of motion for 
$F_{kk'}=\langle\langle c_{{\bf k}\downarrow};c_{{\bf k}'\uparrow}^{\dag}\rangle\rangle_{i\omega_n}$
is also obtained in a similar way.
By using the decoupling procedure for Green's functions [11-13], 
we can obtain a closed solution
for a set of Green's functions.
The Green's function $G_{{{\bf k}}{{\bf k}}'\uparrow}(i\omega_n)$ is obtained as
\begin{equation}
G_{{{\bf k}}{{\bf k}}'\uparrow}(\omega)  =  \delta_{kk'}G_{{\bf k}\uparrow}^0(\omega)- \frac{J}{2N}G_{{\bf k}\uparrow}^0(\omega)\frac{J}{2}
\Gamma(\omega)G_{{\bf k}'\uparrow}^0(\omega)
 \frac{1}{1+JG(\omega)+\left( \frac{J}{2}\right)^2\Gamma(\omega)F(\omega)},
\end{equation}
where the analytic continuation $i\omega_n\rightarrow \omega$ is carried out.
We defined the following functions:
\begin{eqnarray}
G_k^0(\omega) &=& \frac{\omega+\mu}{ (\omega+\mu)^2-v^2(k_x^2+k_y^2)-v_z^2 k_z^2},\\
F(\omega) &=& \frac{1}{N}\sum_kG_k^0(\omega),\\
\Gamma(\omega) &=& \frac{1}{N}\sum_k \left(m_k-\frac{3}{4}\right)G_k^0(\omega),\\
G(\omega) &=& \frac{1}{N}\sum_k \left( n_k-\frac{1}{2}\right) G_k^0(\omega),\\
G_{k\sigma}^0(\omega) &=& \frac{\omega+\mu+\sigma v_z k_z}{ (\omega+\mu)^2-v^2(k_x^2+k_y^2)-v_z^2 k_z^2},
\end{eqnarray}
with $m_k=3\sum_q \langle c_{q\uparrow}^{\dag}c_{k\downarrow}S_{-}\rangle$ and
$n_k=\sum_q \langle c_{q\uparrow}^{\dag}c_{k\uparrow}\rangle$.

\section{Kondo Temperature}

From the Green's function $G_{{{\bf k}}{{\bf k}}'\uparrow}(i\omega_n)$, the 
Kondo temperature $T_{\rm K}$ is
determined from a zero of the denominator in this formula.  We consider
\begin{equation}
1+JG(\omega)=0
\end{equation}
in the limit $\omega\rightarrow 0$ by neglecting higher-order term being 
proportional to $(J/2)^2$.
Let us adopt for simplicity that $v_z=v$ and then the dispersion is
$\xi_k=\pm v\sqrt{k_x^2+k_y^2+k_z^2}-\mu$ in three dimensions.
We neglect the term of the order of $J$ in 
$n_k=\langle c_{k\uparrow}^{\dag}c_{k\uparrow}\rangle$.
The equation $1+JG(0)=0$ results in the Kondo temperature $T_{\rm K}$ given as
\begin{equation}
k_{\rm B}T_{\rm K} = \frac{2e^{\gamma}D}{\pi}\exp\left( -16\pi^2\frac{v^3}{\mu^2}\frac{1}{|J|} \right),
\end{equation}
where $D$ is a cutoff and $\gamma$ is Euler's constant.  
We assumed that $\rho(0)|J|$ is small, $\rho(0)|J|\ll 1$ and  
$k_{\rm B}T_{\rm K}\ll |\mu|$ to 
derive the above formula, where $\rho(0)$ is the density of states at the
Fermi surface.
The result shows that the Kondo effect indeed occurs in a Dirac system.
We can consider the Kondo effect in two dimensions by setting $v_z=0$, and also
the $d$-dimensional case in general.
In $d$ dimensions $T_{\rm K}$ reads 
\begin{equation}
k_{\rm B}T_{\rm K} = \frac{2e^{\gamma}D}{\pi}\exp\left( -\frac{8(2\pi)^d}{\Omega_d}
\left(\frac{v}{|\mu|}\right)^{d-1}\frac{v}{|J|} \right),
\end{equation}
where $\Omega_d$ is the solid angle in $d$-dimensional space, namely, the
area of the $(d-1)$-sphere $S^{d-1}$.

In the limit $|\mu|\rightarrow 0$, the equation $1+JG(0)=0$ has no solution.
Hence, when $|\mu|$ is small, $T_{\rm K}$ is reduced and vanishes.
This measn that, when the Fermi surface is
point like, the Kondo effect and thus the Kondo screening never appears because 
of the weak scattering 
by the localized spin.
When $|\mu|$ is small, $|\mu|\ll D$, $T_{\rm K}$ is proportional to $|\mu|$:
\begin{equation}
k_{\rm B}T_{\rm K} = \frac{4e^{\gamma}}{\pi}|\mu|\exp\left( -\frac{8(2\pi)^d}{\Omega_d}
\left(\frac{v}{|\mu|}\right)^{d-1}\frac{v}{|J|} \right).
\end{equation}

When $\rho(0)|J|$ is large being of order O(1) or more than that, 
$T_{\rm K}$ will show an
algebraic dependence on $|\mu|$.
$T_{\rm K}$ reads, being proportional to $|\mu|$,
\begin{equation}
k_{\rm B}T_{\rm K} \simeq \frac{1}{\pi}\sqrt{\frac{7\zeta(3)}{8}}|\mu|
\sqrt{\rho(0)|J|},
\end{equation}
up to the order of $\sqrt{\rho(0)|J|}$ where we put 
$\rho(0)= \Omega_d/(2\pi)^d\cdot |\mu|^{d-1}/v^d$ and
$\zeta(3)$ is the Riemann zeta function 
at argument 3.

\section{Summary}

We investigated the s-d Hamiltonian with the localized spin interacting 
with the Dirac fermions.
The Green's function method is applied to examine a singularity in Green's 
functions to determine the
Kondo temperature $T_{\rm K}$.  The singularity leads to the formula of $T_{\rm K}$ 
as usual being proportional to $\exp(-{\rm const}/\rho |J|)$ with the density 
of states $\rho$.
The Kondo effect indeed occurs in the band of Dirac fermions and will produce 
singular properties in
physical quantities. 
$T_{\rm K}$ vanishes when the Fermi surface is point like.
The term being proportional to $\sigma_z$ like the magnetic field, which we did 
not have considered in this paper,
is also important in the Kondo effect.
It is also interesting to study the nature of interaction between two magnetic
impurities in Dirac metals.  This issue was investigated intensively in
the original Kondo problem\cite{jay81, jon89, yan91, aff95}.

\end{document}